\documentstyle[prb,aps,epsf,psfig,multicol]{revtex}

\newcommand{\ua}{\uparrow}
\newcommand{\da}{\downarrow}

\begin{document}
\draft

\title{Scaling Study of  the Metal-Insulator                 
        Transition   in one-Dimensional Fermion Systems}

\author{Shi-Jian Gu }
\address{Departamento de F\'{\i}sica, Universidade de \'Evora,
Rua Rom\~ao Ramalho, 59, P-7000-671 \'Evora, Portugal
\\ and 
Zhejiang Institute of Modern Physics, Zhejiang University, China.}

\author{Vitor M. Pereira }
\address{Department of Physics, Boston University, Boston, 
Massachusetts 02215, USA,\\ and Centro de F\'{\i}sica do Porto e 
Departamento de F\'{\i}sica, Faculdade de
Ci\^encias, Universidade do Porto, \\
Rua do Campo Alegre 687, P-4169-007
Porto, Portugal.}

\author{N. M. R. Peres}
\address{Departamento de F\'{\i}sica and  
Centro de F\'{\i}sica da Universidade do Minho, Campus Gualtar, 
P-4700-320 Braga, Portugal.}

\date{\today}

\maketitle

\begin{abstract}
We consider the Ising phase 
of the antiferromagnetic XXZ Heisenberg chain on a finite-size lattice
with $N$ sites.
We compute the large $N$ behavior of the spin stiffness, obtaining
the correlation length $\xi$. We use our
results to discuss the scaling behavior of  metal-insulator
transitions in 1D systems, taking into account the mapping
between the XXZ Heisenberg chain and the spinless fermion
model and known results for the Hubbard model. 
We study the scaling properties
of both the  Hubbard model and the XXZ Heisenberg chain,
by solving numerically the Bethe Ansatz equations.
We find that for some range of  values of $\xi/N$ the scaling
behavior may be observed for the Hubbard model but not for
the XXZ Heisenserg chain. 
We show how $\xi$ can be obtained
from the scaling properties of the spin stiffness for small system sizes.
This method can be applied to models having not an exact solution, 
illuminating their transport properties.
\end{abstract}
\vspace{0.3cm}
\begin{multicols}{2}
\section{Introduction}

Since Bethe's solution \cite{bethe} of the one-dimensional (1D) isotropic 
Heisenberg model, that the Bethe Ansatz (BA)
method for solving and understanding
1D many-body integrable systems has been a valuable tool 
in condensed matter.

Many physical quantities can be computed from the BA 
equations alone. Among these  are thermodynamic quantities,
which can be computed using the thermodynamic Bethe ansatz 
method.\cite{takahashi} 
Nevertheless, the BA solution does not provide a way of computing 
all physical quantities of interest. For example, non-diagonal
correlation functions, such as the spectral function, important to
understand photoemission experiments, or the optical conductivity, important
to understand the response of the system to an electromagnetic field
have to determined using other methods.\cite{prl,luther,carmelo}

The study of transport properties in integrable systems
is one of the most active fields of research in integrable models in these 
days. The two most studied models have been the Hubbard  and the 
spinless fermion (or the spin 1/2 Heisenberg chain) models.
The understanding of the transport properties of these systems is related to
the calculation of the optical conductivity
(or spin conductivity \cite{bigcarmelo}). This quantity is defined as
\begin{equation}
\sigma(\omega)=2\pi D_c\delta(\omega)+\sigma_r(\omega)\,,
\label{sigma}
\end{equation}
where $D_c$ is the charge stiffness and $\sigma_r(\omega)$ is the regular
part of the conductivity. This last quantity is associated with
absorption of radiation at finite $\omega$ values. The charge stiffness
is associated with the existence of a metal insulator transition in 
the system,  driven by the interaction (a Mott-Hubbard transition).
If $D_c$ is finite the system presents 
infinite conductivity (free acceleration),
but if  $D_c=0$ the system is an insulator. 
This simple picture holds
at zero temperature. 

W. Kohn  showed that $D_c$ can be computed
from the ground state energy, $E_0$, of the system, imposing 
twisted boundary conditions, as\cite{kohn}
\begin{equation}
D_c = \left. \frac{N}{2}\frac{\delta^2E
}{\delta \phi^2}\right|_{\phi=0}\,,
\label{stif}
\end{equation}
where $\phi$ is a twisting angle (or, in the 1D case, a flux piercing
the ring).
 
At finite temperature, the picture  above is slightly changed but
Eq. (\ref{sigma}) still holds.\cite{castella} The first question
one can raise is whether at finite temperature the system still
has a finite value of $D_c$ if $D_c$ is finite at zero temperature.
The answer to this  question was first conjectured
by Zotos and Prelov\v{s}ek
using numerical methods \cite{prelovsek}, and put on  firm grounds
by Zotos \cite{zotos} in the context of the XXZ Heisenberg model. 
These authors have shown that integrable systems present ballistic transport
of charge and spin at finite temperature if the charge (or spin) stiffness
is finite at zero temperature. 

These theoretical 
questions about the effect of integrability on transport properties
triggered experimental research on transport of energy and spin in quasi 1D
materials, such as SrCuO$_2$ and Sr$_2$CuO$_3$. The results are controversial.
The experimental study of heat transport by spin excitation of 
 SrCuO$_2$ and Sr$_2$CuO$_3$ materials revealed an anomalous enhanced 
thermal conductivity for temperatures above 40 K,\cite{sologubenko}
and the authors concluded for ballistic transport of energy by the spinons.
The MNR study,\cite{thurber} for zero momentum transfer, in the Sr$_2$CuO$_3$
material was able to probe the uniform spin susceptibility
separately from the staggered $\pm \pi/a$ 
($a$ is the lattice constant) mode. The authors concluded for ballistic
spin transport in the $T=0$ limit and for diffusion-like transport
at finite temperatures (even for $T\ll J$, where $J$ is the 
antiferromagnetic coupling). Both from
an experimental and theoretical\cite{pedro} point of view, the study
of transport properties in ladder spin systems is open to research, and
the questions raised in this work should also be pursued in those
systems.

Although the theoretical picture for the spin transport in the XXZ
Heisenberg model is clear if we consider the system in the parameters
region corresponding to the XY universality class [$\Delta <1$,
see Eq.(\ref{hamilt})], a complete
understanding of the spin stiffness 
(or charge stiffness, in terms of the spinless fermion model)
in the parameters region corresponding
to the Ising university class ($\Delta>1$) is lacking. In the XY universality
class region  the zero temperature spin stiffness is given by
$D_c=J\pi\sin(\mu)/(4\mu (\pi-\mu))$, with $1\leq\mu\leq\pi$,
reaching the value $J/4$ at the isotropic point ($\mu=\pi$).
\cite{shastry}
At finite temperature, and for $1<\mu<\pi$, the spin stiffness was
computed by Zotos,\cite{zotos} although there is some
controversy about its  temperature dependence.
\cite{alvarez}
In the Ising universality class region the zero temperature
spin stiffness is zero, in the thermodynamic
limit, and its value for finite temperatures has not
been computed yet but it was conjectured it should also be zero.
\cite{prelovsek,peres}  
We stress here that the above conjecture is by no means obvious.
For example, in the Hubbard model at half-filling, the zero
temperature charge stiffness is zero, but it was found
to be finite at finite temperatures,\cite{kawakami}
although this result has been criticized by computing
the curvature of levels explicitly in the $U\gg t$ limit.\cite{peres00}

Also the scaling properties  of $D_c$ in the region $\Delta>1$ have not been
studied as were in the $\Delta<1$ region,\cite{nicolas} 
and for the Hubbard model. \cite{stafford}   

In the study of the scaling properties of the Mott-Hubbard insulator    
the existence of a scaling function $Y(\xi/N)$ for the
charge stiffness  was shown, \cite{stafford} 
where $\xi$ is the correlation length and $N$
is the number of sites, and it was conjectured that $Y(\xi/N)$
should be a universal function characterizing other metal-insulator
transitions in 1D systems and, in particular, it
should characterize that transition in the spinless fermion
model. 

The spinless fermion model can be mapped into the spin 1/2 Heisenberg chain
by a Jordan-Wigner transformation. 
\cite{mattis}
Therefore the transition between a spin
conductor (spin currents propagating ballisticaly) and a spin insulator
(existence of spin diffusion), occurring at the isotropic
point, and characterized by the spin stiffness value can be mapped into
a metal-insulator transition, driven by the Coulomb interaction 
among spinless fermions at adjacent sites. In this work we consider
the spin language, but our results are equally valid for spinless fermions.

This work is organized as follows. In the section \ref{asymp} we derive
the asymptotic (large $N$) behavior for the spin stiffness and obtain the
correlation length $\xi$ charactering the insulating state.
In section \ref{num} we present a numerical study of the spin stiffness
of the XXZ Heisenberg chain and the Hubbard model. 
In section \ref{sum} we summarize our results.

\section{Asymptotic behavior}
\label{asymp}

The Hamiltonian of XXZ model reads
\begin{eqnarray}
{\cal H}=-J\sum_{l=1}^N [\frac 1 2( S_l^+ S_{l+1}^-e^{-i\phi/N}
+{\rm H. c.})+\Delta
S_l^z S_{l+1}^z],
\label{hamilt}
\end{eqnarray}
where $N$ is the number of sites, $ S_l^+$, $S_l^-$,  $S_l^z$ are spin
1/2 operators at site $l$, and $\phi$ is a flux piercing the 1D ring.
The connection with spinless fermion model is made by means of
the Jordan-Wigner transformation. In terms of spinless fermion
the Hamiltonian (\ref{hamilt}) can be written as

\begin{equation}
      {\cal H}  = -\frac{J}{2} \sum_{\langle j,i \rangle}
        c_{j}^{\dag }c_{i}  +
        V\sum_{i}(\hat{n}_{i}-\frac 1{2})(\hat{n}_{i+1}-
        \frac 1{2})\,,
\label{spinless}
\end{equation}
where the spinless fermion operators $c_{i}^{\dag }, c_{i}$
obey the usual anti-commutation relations, $\langle i,j \rangle$ means
summation over nearest neighbors,  $\hat{n}_{i}=c_{i}^{\dag }c_{i}$
is the usual local number operator, $\frac{J}{2}$ 
is the hopping integral (normally called $t$ but here called
$\frac{J}{2}$ to stress the important exact correspondence between
this model and the XXZ Heisenberg spin 1/2 chain), 
and $V=J\Delta$ is the 
nearest-neighbor Coulomb repulsion.

The Hamiltonian (\ref {hamilt}) can be solved by the BA, 
\cite{takahashi,cloizeaux,gaudin}  
and the energy eigenvalues 
are given, for $\Delta=\cosh(\gamma)$ ($\gamma>0$), by

\begin{equation}
E = E_0- J \sum_{j=1}^{N_\da} \frac{\sinh^2\gamma}{\cosh
\gamma-\cos(\gamma x_j)}.
\end{equation}

where
$E_0=J\Delta N/4$, $N_\sigma$ ($\sigma=\ua,\da$) 
stands for the number of up and down
spins, and
the numbers $x_j$ are discussed below. 
In the case $N_\ua=N_\da$ the system has an energy gap 
relatively to the ground
state. In terms of spinless fermions the system
is an insulator 
(in the thermodynamic limit). The energy gap is given by
\begin{equation}
E_\Delta = J\sinh\gamma\sum_{n=-\infty}^\infty
\frac{(-1)^n}{\cosh(n\gamma)}\,.
\label{ener}
\end{equation} 
The numbers $x_j$ obey Bethe ansatz equations (with twisted boundary
conditions)

\begin{eqnarray}
N\theta_1(x_j,\gamma)&=&2\pi
I_j+\phi+\sum_{l=1}^{N}\theta_2(x_j-x_l, \gamma).
\end{eqnarray}

where
\begin{eqnarray}
\theta_n(x, \gamma)\;=\;2\tan^{-1}\left[\frac{\tan(\gamma
x/2)}{\tanh(n\gamma/2)} \right]\,.
\end{eqnarray}

The ground state quantum number configuration is:
\begin{eqnarray}
I_j=\left[-\frac{N_\da-1}{2},
-\frac{N_\da-3}{2},\dots,\frac{N_\da-1}{2}\right]
\end{eqnarray}

In the absence of magnetic flux, the ground state density is
given by

\begin{eqnarray}
\rho_\infty(x)&=&\frac{1}{2Q}\left(1+\sum_{n=1}^\infty
\frac{\cos(n\gamma x)}{\cosh(n\gamma)}\right) \nonumber \\
&=&\frac{K}{2\pi Q}\,{\rm dn}\left(\frac{Kx}{Q},k\right).
\label{eq:rho0}
\end{eqnarray}

where $K=K(k)$ is the complete elliptic integral\cite{abramowitz} of the
first kind, dn($y, k$) is a Jacobian elliptic function of modulus $k$,
\cite{abramowitz}
and we have 
\begin{eqnarray}
\frac{K(\sqrt{1-k^2})}{K(k)}=\frac{1}{Q}=\frac{\gamma}{\pi}\,.
\end{eqnarray}
The  ground state density (\ref{eq:rho0}) can be obtained from the
momentum density $p(x)=$am$(Kx/Q,k)$ as $dp(x)/dx=2\pi\rho(x)$, and
where am$(y,k)$ is the amplitude function.

In order to obtain the charge-stiffness large-$N$ asymptotic behavior
we write the energy (\ref{ener}) as 

\begin{eqnarray}
E = E_0 - J\int\frac{\sinh^2\gamma \;\,{\rm dx}}{\cosh
\gamma-\cos(\gamma x)} \sum_{j=1}^{N_\da} \delta(x-x_j).
\label{eint}
\end{eqnarray}
Using the properties of the delta function and 
the Poisson summation formula, Eq. (\ref{eint})
can be written as
\begin{eqnarray}
E=E_0 - JN \sinh^2\gamma \int\frac{\rho_N(x,\phi)\;\, {\rm d}x}{\cosh
\gamma-\cos(\gamma x)}
\times\nonumber\\
\sum_{m=-\infty}^\infty \exp\{im[N
p_N(x)-\phi]\}\,,
\label{epois}
\end{eqnarray}
where $p_N(x_j)=(2\pi I_j+\phi)/N$. For large $N$ the integral
in Eq. (\ref{epois}) is dominated by the $m=0,\pm 1$ terms.
As a consequence the stiffness (\ref{stif}) is a sum of two
terms and reads
\begin{equation}
D_c=D_c^{(m=\pm 1)}+D_c^{(m=0)}\,,
\label{2stif}
\end{equation}
where
\begin{eqnarray}
D_c^{(m=\pm 1)} &=& \left. \frac{N}{2}\frac{\delta^2E^{(m=\pm 1)}}
{\delta \phi^2}\right|_{\phi=0}\nonumber\\
 &= &\frac{\gamma JN
\sinh^2\gamma}{2\pi} \int \frac{\sin [Np_N(x)] \sin (\gamma x)
}{[\cosh \gamma-\cos(\gamma x)]^2} \;{\rm d} x\,,
\label{m1}
\end{eqnarray}
and

\begin{eqnarray}
D_c^{(m=0)}&=&\left. \frac{N}{2}\frac{\delta^2
E^{(m=0)}}{\delta\phi^2}\right|_{\phi=0}\nonumber\\
&=& - \frac J 2  N^2
\int {\rm d}x  \frac {\sinh^2\gamma}{\cosh \gamma-\cos(\gamma x)}\left.
\frac{\delta^2\rho_N(x)}{\delta \phi^2}\right|_{\phi=0}.
\label{m0}
\end{eqnarray}

The first term dominates  Eq. (\ref{2stif}) 
for large values of $\Delta$, since the second one
is, in this case, exponentially small.
Both Eq. (\ref{m1}) and Eq. (\ref{m0}) are evaluated using the
steepest descent method. The final asymptotic result for
$D_c$ can be cast in the form
\begin{equation}
D_c(N) \sim D(\gamma)\sqrt N e^{-N/\xi(\gamma)}\,, 
\label{dassim}
\end{equation} 
where 
$\xi(\gamma)=ip(Q+i)-i\pi N/2$ is the correlation length  given by 
\begin{equation}
1/\xi(\gamma) = \frac{\gamma}{2} + \sum_{n=1}^\infty \frac{(-1)^n}{n}
\frac{\sinh(n\gamma)}{\cosh(n\gamma)}\,.
\label{xi}
\end{equation}
and $D(\gamma)$ is given by
\begin{eqnarray}
D(\gamma)= \left[ \frac{\tanh^2 \gamma}{4}+  \sum_{n=1}^\infty (-1)^{n+1}
n e^{-2n\gamma}\tanh(n\gamma) \right]\nonumber\\
\times\frac{\gamma J \sinh\gamma
}{|2\pi f''(z_0)|^{1/2}}.
\end{eqnarray}

\begin{figure}[htf]
\begin{center}
\epsfxsize=7cm
\epsfbox{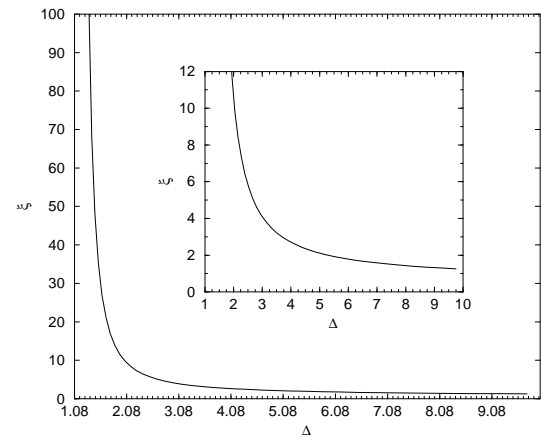}
\end{center}
\caption{Representation of the correlation length 
as function of $\Delta$. It is clear that $\xi$ diverges
as we approach the isotropic point $\Delta=1$.}
\label{xidel}
\end{figure}

The correlation length (\ref{xi}) agrees with the value
obtained previously using a  different method,
\cite{baxter} and is represented in Fig. \ref{xidel}.

 The result (\ref{xi}) is quite similar to that obtained for
the Hubbard model by Stafford, Millis, and Shastry.
\cite{stafford} 
\section{Numerical results}
\label{num}

\begin{figure}[htf]
\begin{center}
\epsfxsize=7cm
\epsfbox{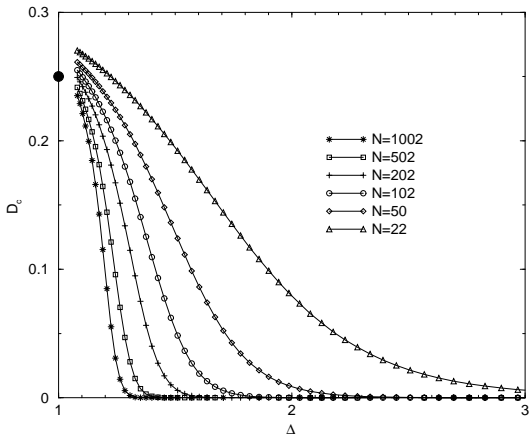}
\end{center}
\caption{Representation of the stiffness, $D_c$, as function of
$\Delta$ for different system sizes. The value of $D_c$ at the isotropic
point ($\Delta=1$)
in the thermodynamic limit is represented by a filled circle.
The represented values of $\Delta$ satisfy the condition
$\Delta>\cosh(0.4)$.}
\label{figDdel}
\end{figure}

At $T=0$ the XXZ Heisenberg model for $\Delta>1$ presents 
non-ideal spin
transport in the thermodynamic limit, because the spin stiffness
is zero. In terms of spinless fermions one has an insulating behavior.
For a finite-size system $D_c$ is finite even for large values of 
$N$ and for moderate values of $\Delta$, as can be seen in 
Fig. \ref{figDdel}.  In this figure 
we see that even for values of $N$ as large as 1000 sites there are
values of $\Delta$ for which $D_c$ is of the order of  its
value at the isotropic point (represented by a filled circle). 
This shows that the behavior of $D_c$, when $\Delta>1$, in  a 
finite-size system, is dominated
by the value of the $D_c$ for $\Delta=1$. The value of $D_c$ at the isotropic
point is by no means trivial, and the presence of this point
has consequences for the scaling behavior of $D_c$ in the Ising
region as we show below.  We believe this difference is due to the
presence of logarithmic corrections to the scaling function by  marginal
interactions at the isotropic point.

\begin{figure}[htf]
\begin{center}
\epsfxsize=7cm
\epsfbox{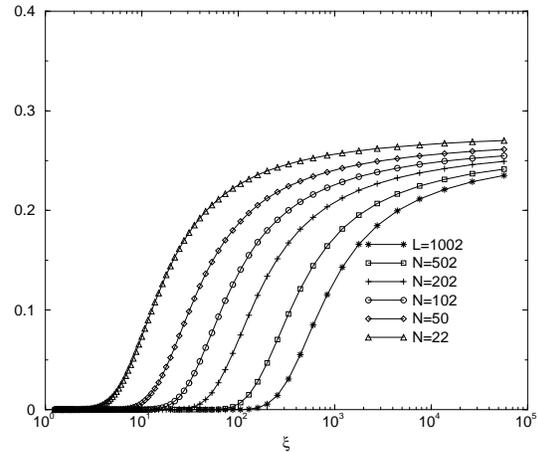}
\end{center}
\caption{Representation of the stiffness, $D_c$, as function of
$\xi$ for different system sizes. The values of $\xi$
correspond to values of $\Delta$ satisfying the condition
$\Delta>\cosh(0.4)$.}
\label{figDxi}
\end{figure}

The above discussion stresses that for mesoscopic systems the 
separation between an insulator and a metal is not clear. Only when
the correlation length is smaller than the system size a clear
separation between an insulator and a metal exist. The correlation
length (\ref{xi}) is independent of the system size, and the
dependence of $D_c$ on $\xi$ is represented in Fig. \ref{figDxi}.
This figure makes the above statement clear, for example, for values
of $\xi$ bigger than $10^2$, $D_c$ is finite even for $N=1002$, whereas
for values of $\xi$ smaller than $10^2$ $D_c$ is zero for 
$N=1002$ but not for the other (smaller) values of $N$.

\begin{figure}[htf]
\begin{center}
\epsfxsize=7cm
\epsfbox{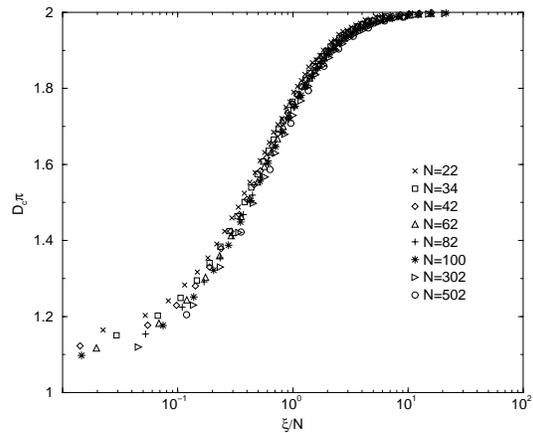}
\end{center}
\caption{Representation of the stiffness, $\pi D_c$, as function of
$\xi/N$ for the Hubbard model, for system sizes
$N=$24, 32, 42, 62, 82, 100, 302 and 502. The correlation
length $\xi$ is given by Eq. (3.2) of Ref. 20.}
\label{hubsca}
\end{figure}

In Ref. \onlinecite{stafford}  
the charge stiffness of the Hubbard model, at half-filling, was shown
to present a scaling behavior
of the form $Y(\xi/N)$, for all values of the Hubbard interaction $U$,
where $\xi$ is the correlation length
and $N$ is the system size. In Figure \ref{hubsca} we show data from a
numerical solution of the BA equations for the Hubbard  model, which 
agrees with the results of Ref.  \onlinecite{stafford}.
Looking more carefully to Figure \ref{hubsca} we see there is scaling
behavior for all values of $N$ as we get closer  to the
gapless phase $U=0$ ($\xi\rightarrow\infty$). On the other hand, as $U$
increases, $\xi\rightarrow 0$, and the scaling behavior can only be verified
for larger system sizes.

Based on the scaling behavior of the Hubbard model, a conjecture
that the same scaling behavior would apply to other 1D systems 
(in particular for the spinless fermion system) presenting
metal-insulator transitions was drawn. 

\begin{figure}[htf]
\begin{center}
\epsfxsize=7cm
\epsfbox{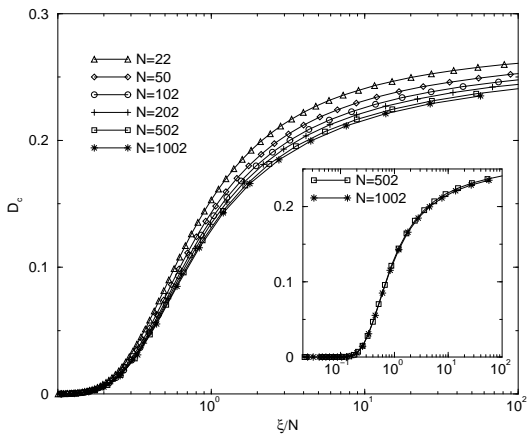}
\end{center}
\caption{Representation of the stiffness, $D_c$, as function of
$\xi/N$, for the Heisenberg model.
The {\bf inset} shows the same but for $N=$502 and 1002, being
clear that the collapsing of the data close to
$\Delta=1$ requires larger values of $N$.}
\label{dcxi}
\end{figure}

Following these ideas  we have first derived
an asymptotic expression for the stiffness of the XXZ Heisenberg model
(or spinless fermions). This allowed us to determine the correlation 
length $\xi$ for this model. We have then solved numerically the BA
equations for systems of different sizes and tried to see if some type
of scaling was possible. From Figure \ref{dcxi} we see that  
the representation of $D_c$ as function of $\xi/N$ gives curves
that do not collapse into each other unless
$N>202$.
This larger value of $N$ need for scaling
to be observed contrast with the case of the Hubbard
model in Figure \ref{hubsca}, although the values of
$\xi/N$ are the same in both cases. 
Also, as we get closer and closer to $\Delta=1$, where
the system presents a gapless phase 
(in Figure \ref{dcxi} this corresponds to $\xi/N\rightarrow\infty$), 
the collapsing of the data
takes place for larger and larger values of $N$. 
This behavior 
contrasts  with that  seen above for the Hubbard model.

\begin{figure}[htf]
\begin{center}
\epsfxsize=7cm
\epsfbox{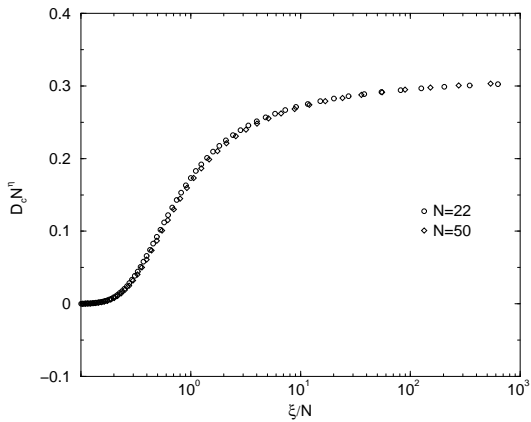}
\end{center}
\caption{Representation of the stiffness, $N^{\eta}D_c$, as function of
$\xi/N$, for $N=22$ and $N=50$.
}
\label{etascaling}
\end{figure}

Since the scaling as function of $\xi/N$
is not perfect for all values of $N$, let us assume that $D_c$
could follow a scaling behavior of the form $D_c=N^{-\eta}f(\xi/N)$ and
let us  find the value of the exponent $\eta$ that leads to the collapsing
of the data.  We proceed by grouping the several $N$ in groups of two
as follows $\{22,50\}$, $\{50,102\}$, ..., $\{502,1002\}$, and
for each set $\eta$ is determined by a minimization procedure
which leads to the best possible scaling. 
Our findings
are listed in Table \ref{group}, and show that the value of the exponent
$\eta$ decreases as the $N$ increases, leading for very large values
of $N$, to the predicted scaling behavior  $D_c=f(\xi/N)$.
On the other hand,
for small values of $N$ the scaling behavior $D_c=N^{-\eta}f(\xi/N)$
works much better as we show in Figure \ref{etascaling}. 
The fact that it was not possible to find single scaling behavior 
for for all $N$ is an indication that  either a 
different scaling law or corrections 
containing  $N$ and associated with the presence of the critical point
are need.

\begin{figure}[h]
\begin{center}
\epsfxsize=7cm
\epsfbox{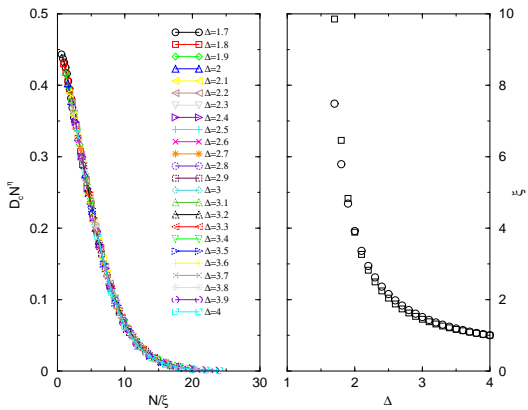}
\end{center}
\caption{On the {\bf left} panel we present the collapse of the numerical
data for several distinct \protect\( \Delta \protect \) according
to the scaling relation (\ref{Eq:2}). On the {\bf right} are our numerical
results for \protect\( \xi \left( \Delta \right) \protect \) obtained
by means of the described method ({\bf squares}), together with 
the exact result ({\bf circles}) obtained
from the Bethe ansatz, for comparison.}
\label{xinum}
\end{figure}

Let us now show that scaling relation $D_c=N^{-\eta}f(\xi/N)$
can be used to determine $\xi$ directly from the solution of 
small systems. 
The scaling relation
\begin{equation}
\label{Eq:1}
g\left( N,\xi \right) =\frac{1}{N^{\eta }}f\left( \frac{\xi }{N}\right) 
\end{equation}
has an interesting property which can be used in order to determine
the correlation length \( \xi  \) in cases where this can not be
done analytically. Supposing that (\ref{Eq:1}) holds for a given quantity,
\( g \), and assuming the existence of a correlation length depending
solely on other parameter of the problem, i. e. \(\xi= \xi (\lambda ) \)
(which, in the case of our problem, is the anisotropy $\Delta$), the scaling
relation can be expressed in a slightly different, but equivalent
notation as
\begin{equation}
\label{Eq:2}
g_{\lambda }\left( N\right) \equiv g\left( N,\xi (\lambda )\right) 
=\frac{1}{N^{\eta }}f\left( \frac{\xi (\lambda )}{N}\right) \, \, .
\end{equation}
We can now consider, as is  frequently the case in numerical calculations, 
measuring \( g \) for various sizes and different values of the parameter
\( \lambda  \). Then, for each pair of sets \( g_{\lambda _{i}}(N) \)
and \( g_{\lambda _{j}}(N) \) we can write
\begin{equation}
\label{Eq:3}
g_{\lambda _{i}}\left( N\frac{\xi (\lambda _{i})}{\xi (\lambda _{j})}
\right) \equiv \frac{1}{N^{\eta }}
\left(\frac{\xi (\lambda _{j})}
{\xi (\lambda _{i})}\right)^\eta
f\left( \frac{\xi (\lambda _{j})}{N}\right) \, \, ,
\end{equation}
or, setting \( r_{ij}=\frac{\xi (\lambda _{i})}{\xi (\lambda _{j})} \),
\begin{equation}
\label{Eq:4}
g_{\lambda _{j}}\left( N\right) =
g_{\lambda _{i}}\left( Nr_{ij}\right) r_{ij}^{\eta }\, \, .
\end{equation}
This relation should hold for given \( r_{ij} \) and \( \eta  \),
that can be extracted by direct minimization
\cite{tese}. Proceeding in this way
for the entire range \( \left\{ \lambda _{1},...,\lambda _{M}\right\}  \)
in which the numerical values of \( g \) are calculated, the dependence
of \( \xi  \) in the parameter \( \lambda  \) can be obtained, except
for a global multiplicative constant. 
An example of the typical result obtained via this procedure
can be seen in Figure \ref{xinum}
(we remark that for this case the
value of $\eta$ is constant for all values of $N$ and $\Delta$
used). 
Here, numerical results for the
Heisenberg model with chains of sizes 6 to 24 were used to calculate
the charge stiffens \( D_{c} \)\( \left( N,\Delta \right)  \). 
We remark that the values of $N$ used here are easily accessible 
to exact diagonalization.
Although
the scaling relation (\ref{Eq:1}) doesn't seem to apply exactly to
this problem, it was found, for \( \Delta  \) not too close to
the \emph{critical} point, that  a good scaling can be obtained as shown in
the left frame of Figure \ref{xinum}, allowing the numerical calculation
of \( \xi \left( \Delta \right)  \), as shown in the right frame
of  Figure \ref{xinum}. As
can be seen, we were able to calculate the correlation length without
any further information other than the numerical results 
for \( D_{c} \)\( \left( N,\Delta \right)  \).
This concept can be usefully applied to other cases where the analytical
result for $\xi$ is unknown, at least to give a rough idea of the behavior
of the correlation length or any other quantity involved in the scaling
relation.

\section{Summary}
\label{sum}

We have studied the spin stiffness of the XXZ Heisenberg chain for $\Delta>1$.
We have shown that the in large $N/\xi$ behavior of the stiffness
agrees with the asymptotic behavior $D(N)=Y(N/\xi)$, where
$Y(x)$ is a universal scaling function, but not
for all ranges of $N/\xi$, as in the Hubbard model. 
Contrary to the case of the Hubbard
model and for a given value of $\xi/N$ the system sizes required
to observe the scaling behavior are about an order of magnitude
larger for the XXZ Heisenberg chain. Moreover as we get closer and
closer to the isotropic ($\Delta=1$) point,
characterized by $\xi/N\rightarrow\infty$, 
larger and larger system sizes are required, indicating
the scaling behavior does not apply in that region.
We believe the difference between the Hubbard model and the
XXZ Heisenberg chain is related to the presence of the 
non-trivial point $\Delta=1$, for the Heisenberg chain as opposed
to the trivial point $U=0$ (independent electrons)
for the Hubbard model. An interesting  question which remains to be 
answered concerns the scaling properties 
and the correlation length of the metal-insulator
transition in an extended Hubbard model, considering
next-nearest neighbor interaction, since this model is 
no longer integrable.  For small system sizes
the above scaling function does not apply and a scaling
of the form $D(N)=N^{-\eta}f(N/\xi)$ produces a
much better scaling behavior. Using this scaling function
the correlation length $\xi$ was obtained from the
analysis of the small systems alone, and shown
to agree well with the exact value. This method
can be used to derive the correlation length for systems
which do not have an exact solution.

This work is supported by  the Portuguese Program
PRAXIS XXI under grant number 2/2.1/FIS/302/94.
V.M.P. is support by Funda\c{c}\~ao para a Ci\^encia e Tecnologia
grant N$^o$ BD/46/55/2001.
We thank P.D. Sacramento for helpful comments on the manuscript.
V.M.P. thanks J.M.B. Lopes dos Santos for valuable discussions about
exact diagonalization methods.
N.M.R.P. thanks X. Zotos for valuable comments on an early stage
of this work.


\begin{table}
\caption{Value of the exponent $\eta$ leading to the collapse
of the data, for all values of $\xi/N$, giving a pair of consecutive
$N$'s.}
\label{group}
\begin{tabular}{cc}
pairs of $N$ & $\eta$ \\  
\hline\\
22 and 50    & 0.040\\
50 and 102   & 0.028\\
102 and 202  & 0.022\\
202 and 502  & 0.016\\
502 and 1002 & 0.010\\ 
\end{tabular}
\end{table}

\end{multicols}

\end{document}